\documentclass[aps,prl,onecolumn,showpacs,groupedaddress, nofootinbib]{revtex4-1}
\usepackage{graphicx}
\usepackage{amsmath,amsthm,amssymb,euscript,mathrsfs,epsf,epsfig}
\usepackage{color}

\def\a{\alpha} \def\b{\beta}  \def\d{\delta} 
   
   \def\l{\lambda} \def\m{\mu}
  \def\p{\pi}  
  \def\t{\tau}  \def\f{\varphi}
   \def\w{\omega}

  \def\P{\Pi}  
   
   \def\W{\Omega}

\def\fr{\frac}  

\begin{document}

\title{Symmetry breaking gives rise to energy spectra of three states of matter}
\author{Dima Bolmatov$^{1}$}
\thanks{d.bolmatov@gmail.com}
\author{Edvard T. Musaev$^{2}$}
\author{K. Trachenko$^{1,3}$}
\address{$^1$ Centre for Condensed Matter and Materials Physics, School of Physics and Astronomy, Queen Mary University of London, Mile End Road, London, E1 4NS, UK}
\address{$^2$ Centre for Research in String Theory, School of Physics and Astronomy, Queen Mary University of London, Mile End Road, London E1 4NS, UK}
\address{$^3$ South East Physics Network}

\begin{abstract}
A fundamental task of statistical physics is to start with a microscopic Hamiltonian, predict the system's statistical properties and compare them with observable data. A notable current fundamental challenge is to tell whether and how an interacting Hamiltonian predicts different energy spectra, including solid, liquid and gas phases. Here, we propose a new idea that enables a unified description of all three states of matter. We introduce a generic form of an interacting phonon Hamiltonian with ground state configurations minimising  the potential. Symmetry breaking $SO(3)$ to $SO(2)$, from the group of rotations in reciprocal space to its subgroup, leads to emergence of energy gaps of shear excitations as a consequence of the Goldstone theorem, and readily results in the emergence of energy spectra of solid, liquid and gas phases. 

\end{abstract}
\pacs{11.10.-z, 11.30.Qc, 63.20.D-}

\maketitle
The main general premise of statistical physics is that observable properties of a macroscopic system can be calculated and explained on the basis of a microscopic Hamiltonian with many degrees of freedom. This has been implemented as a successful program that, notably, has been applied to each of the three states of matter (solids, gases, liquids) {\it individually} \cite{landau}. For example, the model Hamiltonian of a solid enforces oscillations around fixed equilibrium positions \cite{Eistein,Debye}, resulting in the marked restriction on the sampled volume of phase space. On the other hand, a gas state is approached by starting with free particles, switching interactions on and predominantly viewing these as small perturbations. The third state of matter, liquids, occupy an interesting intermediate state with a combination of strong interactions and cohesive state as in solids and large flow-enabling particle displacements as in gases. This combination is believed to preclude the calculation of thermodynamic properties of liquids in general form \cite{landau}.

The general problem represented by liquids is well-known \cite{landau,Born}, yet here we begin with asking an even more fundamental question. The question bears on some deep issues that were recognized long ago \cite{frenkel} yet remain unsolved, those of operating in restricted phase space rather full phase space. As in the example above, most model Hamiltonians of solids impose restrictions on the phase space where atoms never leave their equilibrium sites. Even with anharmonicity of interactions properly introduced, modern statistical physics can not predict whether and under what conditions a given Hamiltonian corresponds to a solid, a liquid or a gas. This is often illustrated as a story of some best physicists who are gathered on an island, given a Hamiltonian and failed to analytically find which state of matter it corresponds to, despite being surrounded by water.

Here, we ask whether a Hamiltonian can be proposed that demonstrably describes energy spectra corresponding to solid, liquid and gas phases. To address this challenge we operate in terms of the phonon Hamiltonian. Ground state configuration breaks the symmetry and the Hamiltonian readily describes energy spectra corresponding to solids, liquids and gases (both interacting and ideal). In this picture, the energy gaps of shear excitations naturally emerge as a consequence of the Goldstone theorem. The group of rotations in reciprocal space $SO(3)$ is spontaneously broken to its subgroup $SO(2)$. Consequently, different choices of couplings of fields correspond to energy spectra of distinct states of matter in reassuring agreement with experimental results.

\begin{figure*}[h]
	\centering
\includegraphics[scale=0.35]{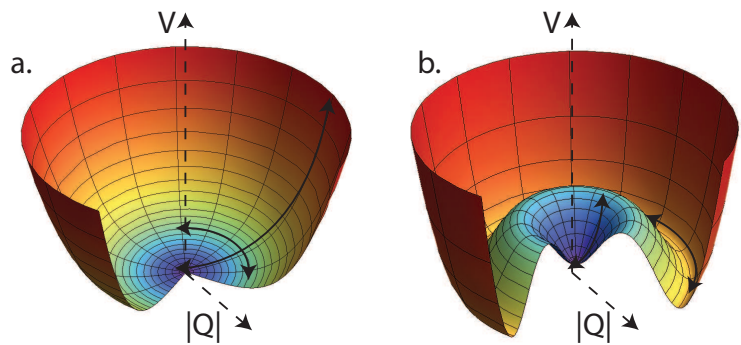}
\caption{When $\w_k>\omega_{\rm F}$, the global minimum is given by $|Q_k^\a|=0$ (a). For $\w_k < \omega_{\rm F}$, the global minimum of the potential is provided by the family of solutions $|{Q}^\a_k|_+$ that breaks the symmetry $SO(3)\to SO(2)$ (b).}
	\label{fig1}
\end{figure*}


\section{Results}
We start with the Hamiltonian describing the dynamics of the phonon field in harmonic approximation \cite{kit63}
\begin{equation}
\label{H_0}
H_0=\fr{1}{2}\sum_{\w_k<\omega_{\rm D}}\left[\P_k^\a\P_{-k}^\a+\m\,\w_k^2 Q_k^\a Q_{-k}^\a\right].
\end{equation}
Here the small Greek indices run from 1 to 3 labelling three space directions and $k$ is a multiindex $\{k_1,k_2,k_3\}$ that denotes the wave vector of the corresponding harmonics and $\omega_{\rm D}$ is the Debye frequency. The parameter $\m$ that takes values 1 or 0 was introduced for further convenience. Summation over the repeated space indices is always assumed and the metric has the signature $\{+,+,+\}$. The collective canonical coordinates $\P^\a_k$ and $Q_k^\a$ are introduced as
\begin{equation}
\begin{aligned}
Q_k^\a &=  \sqrt{m} \sum_{j=1}^N e^{{\texttt i}L j\cdot k }x^\a_j,\\
\P^\a_k &= \dot{Q}_k^\a,
\end{aligned}
\end{equation}
where $x_j^\a$ are 3 coordinates of the $j$-th atom of the lattice, $L$ is the lattice spacing , $N$ is the total number of atoms, ${\texttt i}$ is the imaginary unit  (${\texttt i}^2=-1$) and $m$ is the mass of an atom in the lattice. The coefficient $\w_k^2$ gives dispersion relation of a phonon. Normal modes satisfy $Q_k^\a=Q_{-k}^\a{}^*$ since coordinates of atoms $x^\a_j$ are real, where star denotes the complex conjugation.

The Hamiltonian \eqref{H_0} that is quadratic in fields defines a free theory with no interactions between phonons. To introduce an interaction one adds a term $H_{int}$ that is of higher order in fields which leads to spontaneous symmetry breaking \cite{rubakov2009}.  The simplest possible interaction term which does not involve fractional powers is of the sixth order in fields \footnote{ In general one may add terms of higher orders in fields. However in order to have a metastable configuration and a family of global minima (see Figure 2) one needs terms of powers $2\cdot (2n-1)$ and $2n$ for integer $n>1$.}
\begin{equation}
\begin{aligned}
H_{int}& =\sum_{\w_k<\omega_{\rm D}}\left[-\fr{g}{2} \left(Q_{k}^\a Q_{-k}^\a\right)^{2}+\fr{\l}{6}\left(Q_{k}^\a Q_{-k}^\a\right)^3\right]\\
& =\sum_{\w_k<\omega_{\rm D}}\left[-\fr{g}{2} |Q_{k}^\a|^4+\fr{\l}{6}|Q_{k}^\a|^6\right],
\end{aligned}
\end{equation}
where $g,\l\in \mathbb{R}^+$ are some real non-negative coupling constants (see discussions in conclusions) and $|Q_k^\a|=(Q_{k}^\a Q_{-k}^\a)^{1/2}$. The total Hamiltonian $H=H_0+H_{int}$ is invariant under the following transformations
\begin{equation}
\begin{aligned}
Q_k^\a & \rightarrow R^\a{}_\b Q_k^\b,\\
\P_k^\a & \rightarrow R^\a{}_\b \P_k^\b,
\end{aligned}
\end{equation}
for any $||R^\a{}_\b ||\in SO(3)$.

The configurations $\bar{Q}^\a_k$ and $\bar{\P}_k^\a$ that minimise energy of the system, break the $SO(3)$ symmetry to $SO(2)$ for a certain range of frequency $\w_k$. The kinetic energy is minimal at configurations $\bar{\P}_k^\a=0$ and the minimum of the potential term can be found in the usual way
\begin{equation}
\label{dVdQ}
\fr{\d V[Q_k^\a]}{\d Q^\b_l}=0,
\end{equation}
where the potential $V[Q_k^\a]$ is defined as
\begin{equation}
\label{V}
V[Q_k^\a]=\sum_{\w_k<\omega_{\rm D}}\left[\fr{\m}{2}\,\w_k^2 |Q_k^\a|^2-\fr{g}{2} |Q_{k}^\a|^{4}+\fr{\l}{6}|Q_{k}^\a|^6\right].
\end{equation}
The equation \eqref{dVdQ} is of the fifth order in $|Q_k^\a|$ and therefore has five solutions. We choose only non-negative roots
\begin{equation}
\label{vac}
\begin{aligned}
|{Q}_k^\a|_{\pm}& =\left(\fr{g}{\l}\pm \sqrt{\fr{\omega_{\rm F}^2-\w_k^2}{\l}}\right)^{1/2},\\
|{Q}_k^\a|_{0}&=0,\\
\omega_{\rm F}^2& \equiv \fr{g^2}{\l}.
\end{aligned}
\end{equation}
The factor $\m$ was omitted here since it takes value 1 for non-trivial cases (see next section). The solution \eqref{vac} behave quite differently when $\w_k > \omega_{\rm F}$ and $\w_k < \omega_{\rm F}$.  Namely, for the frequencies $\w_k > \omega_{\rm F}$ all three roots coincide and the potential has only one minimum $|{Q}_k^\a|=0$ that is invariant under the $SO(3)$ transformations. However, for $\w_k < \omega_{\rm F}$ the global minimum of the potential is provided by the family of solutions $|{Q}^\a_k|_+$ that is not invariant under $SO(3)$ and spontaneously breaks the symmetry to $SO(2)$.

When $\w_k<\omega_{\rm F}$ the solution $|Q_k^\a|_{0}=0$  represents the local metastable minimum. Indeed, two roots $|Q_k^\a|_{\pm}$ correspond to two extrema of the potential, one of which (with the minus sign) is a local maximum (see Fig.1). This means that  the pseudo-vacuum state $|Q^\a_{k}|=0$ is stable on the classical level, but becomes metastable if quantum effects are taken into account. This leads to quantum tunneling of the state $|Q_k^\a|=0$ to the true vacuum state given by $|\bar{Q}^\a_k|\equiv|Q_k^\a|_{+}$ and symmetry breaking.

According to the Goldstone theorem this leads to two massless modes $\f_k^{2,3}$, which we call transverse modes, one for each broken symmetry generator, and one massive mode $\f^1_k$, which we call longitudinal mode \cite{goldstone, stroc}.  Hence, the longitudinal mode is the one that corresponds to the unbroken  symmetry generator. Excitations of the phonon field around the ground state $\bar{Q}_k^\a$ can be written as
\begin{equation}
Q^\alpha_k=\bar{Q}_k^\alpha+\f_k^\alpha.
\label{vacuum}
\end{equation}
For a particularly chosen vacuum $\bar{Q}_k^\a=\d^\a_1|\bar{Q}_k|$ we obtain the following Hamiltonian:
\begin{equation}
\label{Ham}
\begin{aligned}
H[\f^\alpha_k]& =\fr{1}{2}\sum_{\w_k<\omega_{\rm D}}\p_k^\alpha\p_{-k}^\alpha+\sum_{\w_k<\omega_{\rm D}}\left[\fr{\W_k^2}{2}\f^1_k\f^1_{-k}\right]+\\
&+ \sum_{\omega_{\rm F}<\w_k<\omega_{\rm D}}\left[\fr{\m\w_k^2}{2}(\f^2_k\f^2_{-k}+\f^3_k\f^3_{-k})\right]+\\
&+V_{int}[\f^\alpha_k]+V_0.
\end{aligned}
\end{equation}
Here $\f_k^\alpha$ are small excitations around the vacuum state and $\p_k^\alpha=\dot{\f}_k^\alpha$ are corresponding canonical momenta. The term $V_{int}$ denotes all higher order interactions and includes all three modes $\f_k^\alpha$ while the last term is an irrelevant shift of the total energy of the system
\begin{equation}
V_0=\sum_{\w_k<\omega_{\rm F}}\left(\fr{\m\w_k^2}{2}|\bar{Q}_k|^2-\fr{g}{2}|\bar{Q}_k|^4+\fr\l6|\bar{Q}_k|^6\right).
\end{equation}
Here the cutoff $\w_k<\omega_{\rm F}$ reflects the fact that for  $\w_k>\omega_{\rm F}$ we have $|\bar{Q}_k|=0$ and the corresponding potential becomes zero.

One should note that the frequency $\W_k^2$ of the longitudinal mode $\f_k^1$ defined as
\begin{equation}
\W_k^2=
\left\{
\begin{aligned}
& 4\left(\omega_{\rm F}^2-\w_k^2\right)+ 4\omega_{\rm F}\sqrt{\omega_{\rm F}^2-\w_k^2}>0,& \w_k< \omega_{\rm F}, \\
& \w_k^2, & \w_k> \omega_{\rm F},
\end{aligned}
\right.
\end{equation}
is non-continuous at the point $\w_k=\omega_{\rm F}$ and $\m$ is again set to 1. Since the symmetry is broken and the system is in the true ground state $|Q_k^\a|\neq0$ the plus sign in \eqref{vac} should be chosen.

At the same time, the Goldstone theorem asserts that the transverse modes $\f_k^2$ and $\f_k^3$ do not contribute to the energy of the system at the quadratic level for $\w_k\leq \omega_{\rm F}$. The term $V_{int}$ that encodes interactions between all three modes involves all frequencies $\w_k\in(0,\w_{D})$. The detailed analysis of the physical consequences of these facts is given in the next section.

Finally, it is worth mentioning that the direction of $\bar{Q}^a_k$ is chosen spontaneously and the form of the resulting Hamiltonian does not depend on this choice.

Identifying the field $\f^1_k$ with longitudinal normal mode and the fields $\f^{2,3}_k$ with transverse shear modes one can write energy of the theory defined by the Hamiltonian \eqref{Ham} as
\begin{equation}
\label{Frenk}
E=K+P_l+P_s(\w_k>\omega_{\rm F})+E_{int}
\end{equation}
\noindent where $K$ is the total kinetic energy and $P_l$ and $P_s$ are the potential energies of longitudinal and shear modes, respectively, and $E_{int}$ corresponds to higher-order terms such as an anharmonicity. The Eq. (\ref{Frenk}) implies that contributions of transverse modes with frequencies $\omega <\omega_F$ to linearised energy vanish. This means, that we do not have free propagating transverse modes with such frequencies.

According to Eq. (\ref{Frenk}), the system supports one longitudinal mode and two shear modes with frequency larger than $\omega_{F}$. Our theory therefore predicts a non-trivial and a non-anticipated effect of the frequency cutoff of shear modes. Remarkably, such a cutoff was earlier discussed on purely dynamic grounds, a point to which we return below. Here, we note that our symmetry breaking approach essentially captures the earlier dynamic idea \cite{frenkel}.

\section{Discussion}
The most intriguing feature of the proposed formalism is albeit the energy \eqref{Frenk} can be interpreted as the energy of a liquid, in fact it describes all three phases of matter depending on the parameters $g,\l$ and $\m$. This is summarized in Table 1.

\begin{table}[ht]
\label{tab}
\centering
\begin{tabular}{p{2cm}p{3.5cm}p{1.5cm}}
\hline
\hline \\[-0.2cm]
Phase & Coupling constants & Normal modes \\[0.5ex]
\hline \\[-1.5ex]
{\begin{tabular}{ll}Ideal \\ Gas\end{tabular}}  &  $ \begin{aligned}
			& \m=0,  & &  g\to 0, \\
            & \l\to0, & & \displaystyle \fr{g^2}{\l}\to 0.
			 \end{aligned} $& \begin{tabular}{l}$|\bar{Q}|=0$\\ no modes \end{tabular} \\[0.6cm]
\hline \\[-0.2cm]
 {\begin{tabular}{ll}Interacting \\ Gas\end{tabular}} \  & $ \begin{aligned}
  			& \m=1,  & &  g\neq 0, \\
              & \l\neq0, & & \displaystyle \fr{g^2}{\l}=\omega_{\rm D}.
  			 \end{aligned} $    & $\begin{aligned} & |\bar{Q}|\neq 0, \\
  			  & \f_{k}^{1} \end{aligned}$ \\[0.6cm]			
\hline \\[-0.2cm]
 Liquid   & $ \begin{aligned}
 			& \m=1,  & &  g\neq 0, \\
             & \l\neq0, & & \displaystyle \fr{g^2}{\l}\neq 0.
 			 \end{aligned} $     & $ \begin{aligned} & |\bar{Q}|\neq 0, \\ &\f_{k}^{1,2,3} \left(\omega^{s,s}_{k} >\frac{1}{\tau}\right) \end{aligned}$  \\[0.6cm]
\hline \\[-0.2cm]
{\begin{tabular}{ll}Solid\end{tabular}}   & $ \begin{aligned}
 			& \m=1,  & &  g\ll\lambda, \\
             & \l\neq0, & & \displaystyle \fr{g^2}{\l}\to 0.
 			 \end{aligned} $    & $ \begin{aligned} & |\bar{Q}|\neq 0, \\ &\f_{k}^{1,2,3} \end{aligned}$ \\[0.6cm]

\end{tabular}
\caption{States of Matter. {\bf Ideal Gas}: no elementary excitations; {\bf Interacting Gas}: only longitudinal excitations $0\leq\omega_{k}^{l}\leq\omega_{\rm D}$; {\bf Liquid}: both longitudinal ($0\leq\omega_{k}^{l}\leq\omega_{\rm D}$) and shear ($\omega_{\rm F}\leq\omega_{k}^{s,s}\leq\omega_{\rm D}$) modes; and {\bf Solid}: all modes are supported ($0\leq\omega_{k}^{l,s,s}\leq\omega_{\rm D}$).}
\end{table}

As follows from the Table I, the parameter $\m$ is used to distinguish the phase of the ideal gas when the potential energy is zero. In contrast, the couplings $g$ and $\l$ are model dependent and can be, for example, derived from the experiment (see discussion in the next section). As summarized in the Table, our theory readily gives rise to the different states of matter as follows:

{\it Ideal gas.} The quartic coupling $g$ and the sextic coupling $\l$ are set to be zero as well as the parameter $\m$. This leaves only the kinetic term in the Hamiltonian. Both longitudinal and transverse modes are non-interacting and massless which corresponds to the ideal gas.

{\it Interacting gas}. The Frenkel frequency $\omega_{\rm F}$ becomes equal to the Debye frequency $\omega_{\rm D}$ which eliminates all transverse shear modes. However, in contrast to the case of the ideal gas the longitudinal mode $\f^1_k$  is massive and has non-zero couplings.

{\it Liquid}. Transverse shear modes $\f^{2,3}_k$ with frequencies $\w_k<\omega_{\rm F}$ do not contribute to the Hamiltonian at the quadratic level while the longitudinal mode $\f^1_k$ does not feel the bound $\w_k=\omega_{\rm F}$ since its couplings are continuous.

Notably, our Eq. (\ref{Frenk}) essentially captures the earlier idea of J. Frenkel that as far as propagating modes are concerned, the only difference between a solid and a liquid is that the liquid does not support shear waves at all frequencies as the solid does, but only those with frequency $\omega_{k}>\omega_{F}=\frac{2\p}{\tau}$ \cite{frenkel}. Here, $\tau$ is liquid relaxation time, the average time between two consecutive atomic jumps in one point in space. With a remarkable physical insight, the argument about the liquid vibrational states was developed as follows. At times shorter than $\tau$, a liquid is a solid, and therefore supports one longitudinal mode and two transverse modes, whereas at times longer than $\tau$, liquid flows and loses its ability to support shear stress, and therefore supports the longitudinal mode only as any elastic medium (in a dense liquid, the wavelength of this mode extends to the shortest wavelength comparable to interatomic separations).  Derived on purely theoretical grounds, this idea was later experimentally confirmed, although with a significant time lag (for review, see, e.g. Ref. \cite{Bolmatov1}). This implies that importantly, a liquid supports propagating shear modes with frequency
$\omega>2\pi/\tau$.

We therefore find that the ability of liquids to support high-frequency shear modes with $\omega_{\rm F}$ as a lower frequency cutoff originates in our general approach based on symmetry breaking. This is an unexpected and a highly non-trivial result. For viscous liquids such as B$_2$O$_3$, the experimental evidence was 
available some time ago \cite{grim}. For low-viscous liquids such as Na and Ga, the experimental evidence came about fairly recently when powerful synchrotron radiation sources started to be deployed that mapped dispersion curves in these systems \cite{monaco1,monaco2}. It is reassuring and gratifying that our proposed general approach captures the experimental findings \cite{Bolmatov1,bolmatov2,bolmatov3,bolmatov4}. 
 
Taking the inverse Fourier transform of both sides of (\ref{vacuum}), we find
\begin{equation}
x^\alpha(t)=\bar{x}^\alpha+\xi^\alpha(t)
\label{jump}
\end{equation}
If we associate $\xi^\alpha(t)$ with oscillations around equilibrium positions and $\bar{x}^\alpha$ with translations, the symmetry breaking $SO(3)\to SO(2)$ acquires a microscopic meaning in real space. Namely, no symmetry breaking takes place in solids where atoms do not jump, giving $\bar{x}^\alpha=0$. In liquids and gases, on the other hand, symmetry breaking is due to particle jumps, i.e. spontaneous translations with amplitudes $\bar{x}^\alpha$. 

Glass has been widely viewed as not a separate state of matter but as a slowly flowing liquid, with relaxation time $\tau$ exceeding observation time. When $\tau$ exceeds experimental time scale, the liquid forms glass \cite{dyre}. Therefore, the glass state in our classification scheme originates when $\tau$ reaches a certain large value.

{\it Solid}. All normal modes are supported and $\omega_{\rm F}$ is equal to zero, reflecting the fact that solids are not able to flow. There is no described symmetry breaking in phonon interactions \cite{Debye}.

The Hamiltonian \eqref{Ham} describes solid, liquid and gas states depending on the choice of coupling constants $g$ and $\l$, that in general may depend on wavenumber $k$. The transverse shear modes $\f^{2,3}_k$ for $\w_k<\omega_{\rm F}$ do not contribute to the Hamiltonian on the quadratic level in the liquid regime. In the coordinate space it may correspond to atomic jumps with characteristic time $\t \sim 1/\omega_{\rm F}$.

\begin{figure*}[h]
	\centering
\includegraphics[scale=0.45]{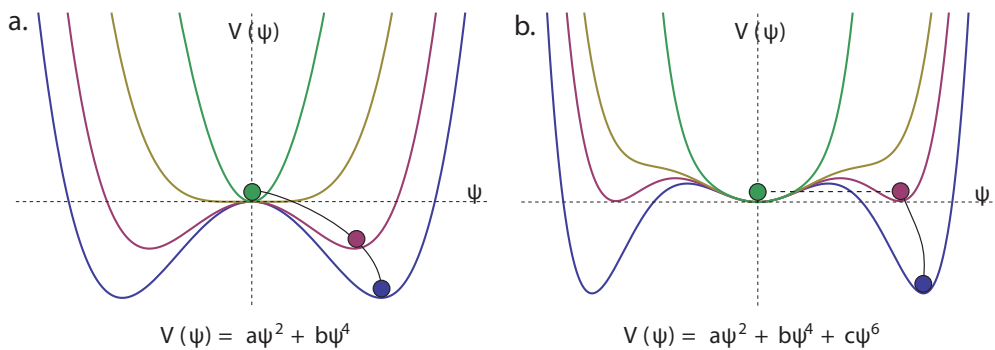}
\caption{Schematic illustration of ground state behaviour for different potentials. Flow of coefficients changes the form of the corresponding potentials and leads to descriptions of different types of phase transitions. The global minimum of the potential on Figure 2.a can be continuously translated to local metastable state and in general describes continuous second-order phase transitions. The translation of the global minimum of the potential on Figure 2.b to local metastable state is discontinuous and in general describes discontinuous first-order phase transitions.}
	\label{s}
\end{figure*}


We now discuss two interesting directions for future work. Identification of the physical meaning of the couplings $g$ and $\l$ from the experiment is an important task. In general these constants may depend on the wavenumber $k$ and have to predict known experimental observables such as, for example, melting and boiling temperature.

The form of the potential in particular can be justified by the following observation. The potential on Figure 2.a in general describes continuous phase transitions. In contrast, the potential on Figure 2.b \cite{Binney}, that was used in the suggested formalism, can be associated with discontinuous phase transitions such as melting (or freezing) . Hence, an intriguing question is a dynamic description of the switch from the liquid to the solid phase regime by investigating the RG flow of the couplings $g$ and $\l$ \cite{Wilson:1973jj}. The description of phase transitions (solid/liquid, liquid/gas and solid/gas) are another remaining challenging tasks. These ideas need more rigorous explanation and connection to the experiment.

The proposed Hamiltonian enables us to describe and predict energy spectra corresponding to other states of matter. To address this challenge one can operate in terms of couplings (see Table I). For instance, from the point of view of energy spectra, plasma does not support transverse modes in the sense of solid state, and therefore falls into the "interacting gas" state in our classification scheme. 

In summery, we have proposed a general form of phonon Hamiltonian with non-trivial minima of potential energy that lead to symmetry breaking. The group of rotations in reciprocal space $SO(3)$ is spontaneously broken to its subgroup $SO(2)$. The energy gaps of shear excitations is a consequence of the Goldstone theorem. Shear modes with $\w_{k}<\w_{F}$ do not contribute to the Hamiltonian at the quadratic level in liquids. It is demonstrated how depending on the couplings $g$ and $\l$, the energy spectra of three basic states of matter (solid, liquid, gas) readily emerge. It is reassuring and gratifying that our proposed general approach captures the experimental findings.

\section{Methods}
In this work we introduce a generic form of an interacting phonon Hamiltonian with ground state configurations minimising the potential. The energy gaps of shear excitations naturally emerge as a consequence of the Goldstone theorem which readily results in the emergence of energy spectra corresponding to solid, liquid and gas phases.


\section{Acknowledgements}
D. Bolmatov thanks Myerscough Bequest and K. Trachenko thanks EPSRC for financial
support. D. Bolmatov acknowledges Thomas Young Centre for Junior Research Fellowship and Cornell University (Neil Ashcroft and Roald Hoffmann) for hospitality.

\section{Contributions}
D. B., E. T. M. and K. T. wrote the main manuscript text and prepared figures 1-2. D. B., E. T. M. and K. T. reviewed the manuscript and have contributed equally to this work.

\section{Competing financial interests}
The authors declare no competing financial interests.

\end{document}